# Deciphering exciton-generation processes in quantum-dot electroluminescence

**Authors:** Yunzhou Deng[1]†, Xing Lin[2]†, Wei Fang[3], Dawei Di[3], Linjun Wang[2], Richard H. Friend[4], Xiaogang Peng[2], Yizheng Jin[1]*

**Affiliations:**

[1]Center for Chemistry of High-Performance & Novel Materials, State Key Laboratory of Silicon Materials, Department of Chemistry, Zhejiang University, Hangzhou 310027, China.

[2]Center for Chemistry of High-Performance & Novel Materials, Department of Chemistry, Zhejiang University, Hangzhou 310027, China.

[3]State Key Laboratory of Modern Optical Instrumentation, College of Optical Science and Engineering; International Research Center for Advanced Photonics, Zhejiang University, Hangzhou 310027, China.

[4]Cavendish Laboratory, University of Cambridge, Cambridge CB3 0HE, UK.

*Correspondence to: Prof. Yizheng Jin (yizhengjin@zju.edu.cn)

†These authors contributed equally to this work.

**Abstract**

Electroluminescence of colloidal nanocrystals promises a new generation of high-performance and solution-processable light-emitting diodes. The operation of nanocrystal-based light-emitting diodes relies on the radiative recombination of electrically-generated excitons. However, a fundamental question—how excitons are electrically generated in individual nanocrystals—remains unanswered. Here, we reveal a nanoscopic mechanism of sequential electron-hole injection for exciton generation in nanocrystal-based electroluminescent devices. To decipher the corresponding elementary processes, we develop electrically-pumped single-nanocrystal spectroscopy. While hole injection into neutral quantum dots is generally considered to be inefficient, we find that the intermediate negatively-charged state of quantum dots triggers confinement-enhanced Coulomb interactions, which simultaneously accelerate hole injection and hinder excessive electron injection. In-situ/operando spectroscopy on state-of-the-art quantum-dot light-emitting diodes demonstrates that exciton generation at the ensemble level is consistent with the charge-confinement-enhanced sequential electron-hole injection mechanism probed at the single-nanocrystal level. Our findings provide a universal mechanism for enhancing charge balance in nanocrystal-based electroluminescent devices.

**Introduction**

Colloidal semiconductor nanocrystals are an important class of solution-processable inorganic luminescent materials[1–5]. In recent decades, remarkable advances have been made in solution-processed light-emitting diodes (LEDs) based on colloidal nanocrystals, including CdSe-based quantum dots (QDs)[6–18], InP-based QDs[19, 20], and perovskite nanocrystals[21, 22], paving the way towards a new generation of display and lighting technologies.

The emissive states for electroluminescence (EL) from QDs are excitons (bound electron-hole pairs). In-depth understanding of the generation and recombination of excitons is essential for interpreting device operation mechanisms and providing guidelines for future developments. Spectroscopic techniques have been applied to study the QD-LEDs[23–25], revealing the recombination dynamics of the in-situ photo-generated excitons. In contrast, few attempts have been made to explore the exciton-generation processes in EL devices based on nanocrystals.

The current understanding of electrical excitation of nanocrystals is largely inherited from the conventional theories developed for bulk semiconductors[14, 17]. A brief picture of this process is that electrons and holes are simultaneously diffused or injected into the emission zone, leading to populations of opposite-sign charge carriers statistically independent of each other[26]. However, distinctive from bulk inorganic semiconductors, nanocrystals show efficient radiative recombination of single-exciton states and even faster non-radiative Auger recombination of charged-exciton states or multi-exciton states because of the strong spatial confinement of charge carriers within individual nanocrystals[27, 28]. In other words, high-efficiency nanocrystal-based EL requires efficient and balanced charge injection to generate single excitons in individual nanocrystals. Thus, it is of fundamental interest to go beyond the conventional "bulk interpretations" and elucidate the unique mechanism for exciton generation in nanocrystal-based EL devices.

Here we address a fundamental question of how the excitons are electrically generated in individual QDs. We develop room-temperature electrically-pumped single-nanocrystal spectroscopy to reveal the hidden elementary processes and the associated kinetics at the single-nanocrystal level. The observations reveal a sequential electron-hole injection mechanism which is enhanced by charge confinement, an intrinsic property of nanocrystals. The unique nanoscopic mechanism is invoked to interpret efficient exciton generation in state-of-the-art QD-LEDs.

**Results**

**Electrically-pumped single-nanocrystal spectroscopy**

We start with the electrical generation of excitons in an individual QD. An isolated CdSe-CdZnS core-shell QD (Supplementary Figure 1) is used as an emitter in a single-dot EL device[29]. In this single-dot device (Supplementary Figure 2), exciton migration and charge transport between neighbouring QDs[30–32] are absent, providing an ideal system for investigating charge dynamics of exciton generation. The single CdSe-CdZnS QD in the EL device exhibits ideal photoluminescence (PL) properties, including single-channel radiative decay and stable single-exciton emission (Supplementary Figure 3). When a forward DC bias of 1.85 to 2.30 V is applied, the single-QD EL device demonstrates background-free and stable single-exciton EL with emission rates of $10^3$–$10^5$ s$^{-1}$ (Supplementary Figure 4).

We develop electrically-pumped single-nanocrystal spectroscopy to probe the elementary charge-injection steps for the exciton generation in the single QD (Fig. 1a). At the single-dot level, the generation of a single exciton may consist of two possible pathways, i.e., the injection of one electron followed by the injection of one hole (QD $\xrightarrow{+e}$ QD$^-$ $\xrightarrow{+h}$ QD$^X$), and the injection of one hole followed by the injection of one electron (QD $\xrightarrow{+h}$ QD$^+$ $\xrightarrow{+e}$ QD$^X$). The intermediate states of the two pathways are different, with one being negatively-charged state (QD$^-$), and another one being positively-charged state (QD$^+$). The charging status of the CdSe-based QDs can be optically distinguished[33, 34]. For the CdSe-CdZnS QD in the single-dot EL device, the lifetimes of the excited states of a neutral QD, a negatively-charged QD, and a positively-charged QD (denoted as X, X$^-$ and X$^+$, respectively) are determined to be 17.4 ± 2.6 ns, 2.7 ± 0.3 ns, and 1.6 ± 0.2 ns, respectively (Supplementary Figure 5). The recombination processes of all optically excited states are much faster than the electrical generation of a single exciton (10–$10^3$ μs, estimated from the emission rates of the single-dot EL device in the given voltage range). Therefore, we use a pulsed laser to excite the single QD under constant electrical pumping, allowing the probing of all possible states relevant to the single-dot EL (Fig. 1b). Note that the pulsed-laser excitation with a low power density (< 0.1 excitation per pulse, 2.5 MHz) does not alter the charging status of the single QD, as demonstrated by the nonblinking PL intensity-time trace[35, 36] (Supplementary Figure 3). In addition, the probability of the electrical injection of a charge carrier into an optically-excited state is extremely low (< 0.2%) because the lifetime of optically excited states in the single QD is very short relative to the time required for electrical excitation (see Methods for details). Hence, the optical excitation serves as a non-invasive probe to monitor the single-dot EL processes.

**Electrical generation of single excitons in a single QD**

A single QD is subject to three excitation conditions, i.e., pulsed-laser excitation, electrical excitation at a constant bias of 2.1 V, and simultaneous electrical and pulsed-laser excitations (Fig. 2a and Fig. 2b). The corresponding PL, EL, and EL-PL intensity-time traces (Fig. 2a) and time-correlated single-photon counting (TCSPC) results (Fig. 2c) are recorded. Comparing with the stable PL and EL intensity-time traces (red and grey regions, Fig. 2a), intensity fluctuations occur in the EL-PL emission (blue region, Fig. 2a). This feature is attributed to electrical excitation-induced fast switching between charged states of the single QD within the bin time (50 ms)[37]. Regarding the TCSPC data of the EL-PL emission (blue curve in Fig. 2c), the EL contribution can be treated as the baseline because the continuous electrical excitation is independent of the pulsed optical excitation. The intensity of EL contribution in the EL-PL emission is identical to that of the EL emission (grey curve in Fig. 2c), confirming the negligible impact of optical excitation on the single-dot EL processes. This feature allows the extraction of the PL decay characteristics of the QD under electrical excitation from the EL-PL emission by subtracting the EL baseline (see Methods for details). The resultant curve (blue, Fig. 2d) is well-described by a double-exponential function. According to the characteristic decay dynamics (Supplementary Figure 5), the fast component (2.7 ns) is assigned as $X^-$, but not $X^+$. The slow component (19 ns) is identical to the single-exciton decay measured without electrical injection (red curve in Fig. 2d). The fact that $X^-$ contributes to the EL-PL emission is supported by spectral analyses. For the CdSe-CdZnS QD, the emission spectrum of $X^-$ is featured by a red-shift of 16 meV compared with that of X (Supplementary Figure 6). As shown in Fig. 2e, the EL-PL spectrum shows broadening in the linewidth and red-shift in the peak wavelength with respect to the spectrum of single-exciton emission. Therefore, the PL response of the QD under electrical excitation is composed of both X and $X^-$, indicating that the exciton generation is mediated by a $QD^-$ state.

Parallel measurements on 15 different single dots by electrically-pumped single-nanocrystal spectroscopy (constant bias: 2.1 V) reveal the exclusive occurrences of $X^-$, ruling out the generation of $X^+$ (Supplementary Figure 7). Furthermore, electrical excitation-dependent experiments show that only X and $X^-$ contribute to the PL response of the single QD in the operating device with EL emission rates in the range of 21,000–70,000 $s^{-1}$ (Supplementary Figure 8). These results unambiguously demonstrate the exclusive existence of $QD^-$ (rather than $QD^+$) as the intermediate state, indicating that the generation of a single exciton follows the pathway of $QD \xrightarrow{+e} QD^- \xrightarrow{+h} QD^X$.

The spectroscopic results are used to analyse the dynamics of the elementary charge-injection steps (Fig. 3). When the single-dot EL device is biased at 2.1 V, the photon emission rate is estimated to be $1.33 \times 10^4$ $s^{-1}$, corresponding to a single-dot EL cycle of ~75 μs ( QD $\xrightarrow{+e} QD^- \xrightarrow{+h} QD^X \rightarrow QD$). Given that the relative luminescent efficiency of $X^-$ to X is ~25% (see the PL blinking trace shown in Supplementary Figure 5), analyses on the PL decay properties of the EL-PL emission (Fig. 2d) reveals that the occupation probabilities of the ground state and the QD$^-$ state in the EL cycle are ~65% and ~35%, respectively (see Methods for details). Accordingly, residence times of the ground state and the QD$^-$ state in the EL cycle are determined to be ~49 μs and ~26 μs, respectively. The results suggest a long-lived intermediate state of QD$^-$. The residence time of the QD$^X$ state, i.e., the single-exciton lifetime (~19 ns), is orders of magnitude shorter than the period of an EL cycle, accounting for the absence of PL responses from the electrically generated QD$^X$ (bi-exciton emission) in the EL-PL emission. The rate coefficients for the injection of one electron into the neutral QD ($k_e$) and for the injection of one hole into the negatively charged QD ($k_h^-$) are derived by using rate equations in the steady-state condition (see Methods for details). The results indicate that $k_h^-$ (3.8×10$^4$ s$^{-1}$) is comparable to $k_e$ (2.0×10$^4$ s$^{-1}$). Considering that no intermediate state of QD$^X$ is detected, the rate coefficient for the injection of one hole into the neutral QD (QD $\xrightarrow{+h}$ QD$^+$), $k_h$, is negligible. Therefore, the rate coefficients of the electrical generation of a single exciton in the single-dot EL device (at a constant bias of 2.1 V) follow $k_h \ll k_e < k_h^-$.

Charge dynamics of the single-dot EL cycle summarized in Fig. 3 suggests that confinement-enhanced Coulomb effects enable the generation of single excitons. In our single-dot EL device, the energy-level alignment of the oxide electron-transport layer, the neutral QD, and the polymeric hole-transport layer favours electron injection and hinders hole injection into a neutral QD (Supplementary Figure 9), leading to $k_h \ll k_e$. The injection of one electron into a neutral QD results in a long-lived state of QD$^-$. For CdSe-CdZnS QDs, the charge carriers are confined within the inner part of the QD (core radius: ~1.6 nm). The small self-capacitance of the nanocrystal induces Coulomb charging effects in the single QD[38–40]. From an energetic point of view, the addition of one electron shifts the electrical potential of the single QD toward higher energies (Supplementary Figure 9). Hence, the rate for injection of a second charge carrier into the negatively charged QD is modulated by confinement-enhanced Coulomb interactions. Specifically, hole injection is significantly enhanced due to Coulomb attractive interaction, resulting in $k_\mathrm{h}^-$ substantially greater than $k_\mathrm{h}$. In consequence, single excitons can be efficiently generated at operating voltages close to the voltage corresponding to the optical bandgap of the QD (Supplementary Figure 4). Furthermore, the fact that $\mathrm{QD}^\mathrm{X}$ is the dominating emissive state for the single-dot EL indicates injection of one electron and one hole into the QD in almost all single-dot EL cycles. Injection of a second electron into the negatively charged QD ($\mathrm{QD}^- \xrightarrow{+\,e} \mathrm{QD}^{2-}$) is greatly suppressed due to repulsive Coulomb interaction.

**Electrical generation of single excitons in QD-LEDs**

We propose that the mechanism of charge confinement-enhanced sequential electron-hole injection derived from EL of an individual QD can be invoked to interpret the exciton generation in the state-of-the-art QD-LEDs (Fig. 4a). We fabricate a red QD-LED comprised of an emissive film of the CdSe-CdZnS QDs (~2 monolayers) and similar charge-transport layers as in the single-dot EL device (Fig. 4a). This device exhibits a low turn-on voltage of 1.7 V (2.4 cd m$^{-2}$) and a peak external quantum efficiency (EQE) of 20.3%, corresponding to an internal quantum efficiency of ~81% (Fig. 4b). These characteristics indicate the efficient generation of single excitons in the QD-LED. At the same time, a large energy barrier for hole injection into the neutral QDs is recognized, as reflected by the fact that the current densities of a hole-only device are approximately three orders of magnitude smaller than those of an electron-only device (Supplementary Figure 10). For this QD-LED with asymmetric energy barriers for charge injection, a conventional scenario of exciton generation via independent charge injection (QD $\xrightarrow{+e}$ QD$^-$; QD $\xrightarrow{+h}$ QD$^+$) followed by a bi-molecular process (QD$^+$ + QD$^-$ → QD$^X$ + QD) would predict unbalanced charge injection and thereby low efficiency, which is in contradiction to the device performance. Instead, we suggest that hole injection is assisted by confinement-enhanced Coulomb interactions. In other words, exciton generation in QD-LEDs relies on the charging of individual QDs (close to the hole-transport layer/emissive layer interface) with electrons, which in turn allows hole injection into the negatively charged QDs (Fig. 4a).

Our hypothesis is verified by in-situ/operando optical measurements on the working QD-LED. TCSPC measurements on the QD film under electrical injection (Supplementary Figure 11 for the experimental setup) show that an additional fast decay component becomes more apparent with increasing electrical-excitation levels, while the slow decay characteristics remain unchanged (Fig. 4c). Estimated from multi-exponential fitting of the curves, the lifetime of the emerged fast component (~2.6 ns) is distinctively different from the characteristic lifetime of $X^+$ and resembles the characteristic lifetime of $X^-$ acquired from single-dot spectroscopy, suggesting the occurrence of QD$^-$ states. The fact that some of the QDs in the working device are negatively charged (with electrons) is also supported by the in-situ measurements of relative PL intensity of the QD layer (see Supplementary Figure 12 for the experimental setup). The results show that the PL efficiency of the QD layer decreases with the increase of current density (Fig. 4d). In the meantime, a significant discrepancy between the changes of PL and EL efficiencies at different current densities is observed. For example, when the current density increases from 0.4 mA cm$^{-2}$ to 3.2 mA cm$^{-2}$ (shaded region in Fig. 4d), the relative PL intensity of the QD film shows a monotonic decrease of up to ~20% while the EQEs (in the range of 19.6% to 20.3%) exhibit minimal relative changes of ~3.4%. In this working regime, electric-field induced effects, which would simultaneously reduce the EL and PL efficiencies[24, 41], are not the main causes responsible for the decrease of PL efficiency. The decrease of PL efficiency of the QD layer can be understood as a portion of QDs is charged with one electron, leading to QD$^-$ states with a lower PL efficiency (~25%). The emergence of QD$^-$ states does not necessarily cause nonradiative Auger recombination in the EL processes[23]. Instead, the negatively-charged individual QDs favour hole injection accelerated by confinement-enhanced Coulomb interactions, allowing the efficient generation of neutral single excitons.

**Discussion**

Our study unravels a nanoscopic picture of sequential electron-hole injection into individual CdSe-based QDs, which is enabled by the confinement-enhanced Coulomb effects, for the efficient generation of single excitons in QD-based EL devices. It is indeed interesting to see that at the single-nanocrystal level, despite that hole injection into a neutral QD is inefficient, the addition of an electron into the QD substantially increases the rate coefficient of hole injection to a level comparable to that of the efficient electron injection. The identification of the long-lived intermediate $QD^-$ state for exciton generation shall provide a new lead on the mechanisms of efficiency loss and device degradation in state-of-the-art CdSe-based QD-LEDs. Furthermore, the electrically-pumped single-nanocrystal spectroscopy developed in this work can be extended for the investigation of electrical excitation processes in other types of nanocrystals, such as InP-based QDs and perovskite nanocrystals.

Considering that Coulomb interactions are pronounced in all nanocrystals with strong carrier confinement, we anticipate the charge confinement-enhanced efficient exciton generation to be a general mechanism for the nanocrystal-based EL devices, including LEDs, electrically-driven single-photon sources[29] and potentially lasers[42, 43]. It would be possible to tailor the charge-confinement effects by engineering the size and dielectric properties of individual nanocrystals. Thus, our findings may lead to new approaches for modulating charge balance in nanocrystal-based EL devices.

**Methods**

**Materials**

Poly (methyl methacrylate) (PMMA, average molecular weight, ~120,000 g mol$^{-1}$) and zinc acetate hydrate (> 98%) was purchased from Sigma Aldrich. Poly (N,N′-bis(4-butylphenyl)-N,N′-bis(phenyl)-benzidine) (poly-TPD, average molecular weight, ~55,000 g mol$^{-1}$) and poly(9,9-dioctylfluorene-co-N-(4-(3-methylpropyl))diphenylamine) (TFB, average molecular weight, ~50,000 g mol$^{-1}$) was purchased from American Dye Source. Colloidal CdSe-CdZnS core-shell red QDs were purchased from Najing technology Co., Ltd. Tetramethylammonium hydroxide (TMAH, 98%) was purchased from Alfa-Aesar. Chlorobenzene (extra dry, 99.8%), octane (extra dry, > 99%) and ethanol (extra dry, 99.5%) were purchased from Acros. Dimethyl sulphoxide (DMSO, HPLC grade) and ethyl acetate (HPLC grade) were purchased from J&K Chemical Ltd. ITO. Colloidal Zn$_{0.9}$Mg$_{0.1}$O nanocrystals were synthesized according to a previous report[44].

**Device fabrication**

The structure of the single-dot EL device is ITO (100 nm)/PEDOT:PSS (40 nm)/poly-TPD (30 nm)/single QDs covered by a PMMA layer (12 nm)/Zn$_{0.9}$Mg$_{0.1}$O (65 nm)/Al (100 nm). Single-dot EL devices were fabricated by depositing materials onto ITO coated glass slides (thickness: ~0.18 mm, resistance: ~50 Ω sq$^{-1}$). PEDOT:PSS solutions (Baytron PVP Al 4083) were spin-coated onto the substrates at 3,500 r.p.m. for 45 s and baked at 150 °C for 30 min. The PEDOT:PSS-coated substrates were subjected to an oxygen plasma for 4 min and then transferred to a nitrogen-filled glove box (O$_2$ <1 ppm, H$_2$O <1 ppm) for subsequent processes. Poly-TPD solutions (in chlorobenzene, 8 mg mL$^{-1}$) were spin-coated at 2,000 r.p.m. for 45 s and baked at 150 °C for 30 min. QD solutions (in octane, ~15 mg mL$^{-1}$, diluted 20,000-folds before use),

PMMA solutions (in acetone, 1.5 mg mL$^{-1}$) and Zn$_{0.9}$Mg$_{0.1}$O nanocrystals (in ethanol, ~30 mg mL$^{-1}$) were layer-by-layer spin-coated onto the substrates at 2,000 r.p.m. for 45 s. Next, Al electrodes (100 nm) were deposited by a thermal evaporation system (Trovato 300C) under high vacuum (~2×10$^{-7}$ torr).

The structure of a QD-LED is ITO (100 nm)/PEDOT:PSS (40 nm)/ TFB (45 nm)/QD (25 nm)/Zn$_{0.9}$Mg$_{0.1}$O (65 nm)/Ag (100 nm). QD-LEDs were fabricated by following procedures similar to those for the fabrication of single-dot EL devices. PEDOT:PSS solutions, TFB solutions (in chlorobenzene, 12 mg mL$^{-1}$), QD solutions (in octane, ~15 mg mL$^{-1}$) and Zn$_{0.9}$Mg$_{0.1}$O nanocrystals were spin-coated onto the ITO coated glass substrates in sequence, and then Ag electrodes were deposited. Film-deposition and post-processing parameters were the same as those for the fabrication of single-dot EL devices, except that the QD solutions were not further diluted for use.

The EL devices were encapsulated by using cover glass slides in a glove box.

**Characterizations of single QDs**

Optical measurements of single QDs were conducted on a home-built fluorescence microscopic system equipped with a 63× oil immersion objective (N.A.=1.46) at room temperature (22–24 °C). The sample holder was a piezoelectric XYZ stage, allowing precise position control. A direct-voltage source (Keithley 2400) was used for electrical excitation and a laser diode (PicoQuant LHD-450) was used for optical excitation. A band-pass filter (621–643 nm, Semrock) was used to block the light from laser excitation. Microscopic images of emissions from single QDs were recorded by an EMCCD (Andor iXon3) with a 100 ms exposure time. Photons from a single QD were directed through a single-mode fibre pinhole and detected by single-photon avalanche photodetectors (SPCMs, PerkinElmer). Intensity traces and TCSPC data of a single QD were simultaneously recorded by a single-photon counting module (PicoHarp 300) operating in the time-tagged time-resolved (TTTR) mode.

For electrically-pumped single-nanocrystal spectroscopy measurements, the laser excitation was set at a low power density (~0.8 W cm$^{-2}$, repetition rate: 2.5 MHz). During each measurement (Fig. 2a), a direct bias of 1.5 V (which does not generate EL), or 2.1 V (which generates EL), is applied to the device. The pulsed-laser excitation is off during the first 15 s and then turned on in the next 15 s. For optical excitation, the occupation probability of the optically-excited states, i.e., the probability of finding the single QD at optically-excited states, can be estimated from the product of the optical-excitation rate and the residence time of the optically-excited state. Given the near-unity PLQY of the CdSe-CdZnS QDs, the optical-excitation rates can be estimated according to the corresponding PL emission rates. In our experiments, the optical-excitation rate is ~1×10$^5$ s$^{-1}$ and the lifetimes of all possible optically-excited states of the QD are shorter than 20 ns, resulting in the occupation probability of the optically-excited states less than 0.2%. Therefore, the probability of the electrical injection of a charge carrier into an optically-excited state is less than 0.2%.

For measurements on single QDs in the PMMA matrix (Supplementary Figure 5), the excitation level was mildly increased (~1.2 W cm$^{-2}$, repetition rate: 1 MHz) to induce the dim states.

Second-order correlation functions of EL or PL from isolated QDs were measured by a Hanbury-Brown and Twiss (HBT) setup consisting of a 50:50 beam splitter. Temporal intensities and time intervals of photons were simultaneously recorded by the single-photon counting module, enabling the extraction of state-resolved second-order correlation functions.

**Characterizations of QD-LEDs**

QD-LEDs were characterized under ambient conditions (room temperature: 22–24 °C and relative humidity: 40–60%). The current density-luminance-voltage (J-L-V) characteristics of the QD-LEDs were measured on a system consisting of a Keithley 2400 source meter and an integration sphere (Ocean Optics FOIS-1) coupled with a spectrometer (Ocean Optics QE-Pro). The devices were swept from zero bias to forward bias. The measurements were repeated many times with the peak EQEs and the efficiency roll-off curves unchanged.

In-situ/operando measurements of transient PL of the QD film in the working QD-LED were conducted in the microscopic system shown in Supplementary Figure 11. A 60× water immersion objective (N.A.=1.0) was used. The power density of the pulsed laser (450 nm, 2.5 MHz) was set to be ~8 mW cm$^{-2}$, which corresponded to an optical excitation level of ~$10^{-3}$ excitation per pulse for each QD. Electrical excitation was limited to a current density smaller than 5.6 mA cm$^{-2}$, corresponding to an electrical injection rate < $3\times10^4$ s$^{-1}$ (~30 μs per electrical excitation) for each QD. The experimental conditions ensured that the excitation levels were controlled in the single-exciton regime when optical excitation and electrical excitation were simultaneously applied. An extra neutral density filter was placed in front of the single-mode fibre. The photon-counting rate is two orders of magnitude smaller than the repetition rate of the pulsed laser, so that the requirement of TCSPC measurement is satisfied[45].

For the measurements of relative PL intensity of QDs in operating QD-LEDs, a continuous 405 nm laser was used for optical excitation, and a direct-voltage source (Keithley 2400) was used for electrical excitation. The laser light was modulated at 1,003 Hz by a chopper and focused on the operating device. The overall emission comprised of EL and modulated PL is collected by a photodetector (Thorlabs PDA100A). Signals from the photodetector were sent to a lock-in amplifier (SR830), where amplitudes of the ac-component of 1,003 Hz, i.e. relative PL intensities, were extracted as the output. The power density of excitation light was kept at a low level to minimize influences of optical-excitation on QD-LED operation.

**Other Characterizations**

The thickness measurements were conducted on atomic force microscopy (Asylum Research Cypher-S) in the tapping mode. The absorption spectra of QD solutions were measured by using a UV-Vis-NIR spectrophotometer (Agilent Cary-5000). The PL spectra were measured by a fluorescence spectrometer (Edinburgh Instruments FLS920). Transmission electron microscope images were obtained by using Hitachi 7700 operated at 80 keV.

**Transient PL analyses**

PL decay curves are fitted with double-exponential functions written as

$$I_t = A_{fast} \cdot e^{-t/\tau_{fast}} + A_{slow} \cdot e^{-t/\tau_{slow}} + B \tag{1}$$

where $I_t$ is the transient light intensity. $\tau_{fast}$ and $\tau_{slow}$ are lifetimes of the two components. $B$ represents background counts, which may include EL emissions. For all normalized PL decay curves, EL emissions are subtracted from the TCSPC data according to the fitting results. The fractional contribution of each component ($f_i$) is given by[45]:

$$f_i = \frac{A_i \cdot \tau_i}{\sum A_i \cdot \tau_i}, \text{ and } f_{fast} + f_{slow} = 1 \tag{2}$$

**Determination of the occupation probabilities**

For a single QD switching between the neutral ground state ("QD" state) and the negatively-charged state ("QD$^-$" state), fractional contributions of the two components in overall PL responses are connected to occupation probabilities of the corresponding states (denoted as $P_{QD}$ and $P_{QD^-}$, respectively) by the following equations:

$$f_{slow} = \frac{P_{QD} \cdot QY_X}{P_{QD} \cdot QY_X + P_{QD^-} \cdot QY_{X^-}} \tag{3}$$

$$f_{fast} = \frac{P_{QD^-} \cdot QY_{X^-}}{P_{QD} \cdot QY_X + P_{QD^-} \cdot QY_{X^-}} \tag{4}$$

where $QY_X$ and $QY_{X^-}$ are the PL QYs of X and X$^-$, respectively. According to Supplementary Figure 5, the relative PL QY of X$^-$, i.e., $QY_{X^-}/QY_X$, is estimated to be 25%.

Given that $P_{QD} + P_{QD^-} \approx 1$, occupation probabilities of the two states are derived from transient PL results:

$$P_{QD} = \frac{f_{slow}}{\frac{QY_X}{QY_{X^-}} \cdot f_{fast} + f_{slow}} \tag{5}$$

$$P_{QD^-} = \frac{f_{fast}}{f_{fast} + \frac{QY_{X^-}}{QY_X} \cdot f_{slow}} \tag{6}$$

**Rate-equation analyses on the single-dot EL cycle**

The single-dot EL cycle (QD $\xrightarrow{+e}$ QD$^-$ $\xrightarrow{+h}$ QD$^X$ $\to$ QD) can be described by rate equations associated with the transitions between the three states. The transition rates for the elementary steps are products of the corresponding rate coefficients ($k_i$) and the occupation probabilities of reactants ($P_i$). The rate coefficients for injection of one electron into the neutral QD (QD $\xrightarrow{+\text{e}}$ QD$^-$) and for injection of one hole into the negatively-charged QD (QD$^-$ $\xrightarrow{+\text{h}}$ QD$^+$) are $k_\text{e}$ and $k_\text{h}^-$, respectively. The rate coefficient of single-exciton recombination (QD$^\text{X}$ → QD) is $k_{\text{X,r}}$, which is the reciprocal of the single-exciton PL lifetime. We connect the occupation probabilities of the three states by the following rate equations:

$$\frac{dP_\text{QD}}{dt} = -k_\text{e} \cdot P_\text{QD} + k_{\text{X,r}} \cdot P_{\text{QD}^\text{X}} \tag{7}$$

$$\frac{dP_{\text{QD}^-}}{dt} = k_\text{e} \cdot P_\text{QD} - k_\text{h}^- \cdot P_{\text{QD}^-} \tag{8}$$

$$\frac{dP_{\text{QD}^\text{X}}}{dt} = k_\text{h}^- \cdot P_{\text{QD}^-} - k_{\text{X,r}} \cdot P_{\text{QD}^\text{X}} \tag{9}$$

For the CdSe-CdZnS QDs with a near-unity PL QY, the photon emission rate of a single-dot EL device ($I_\text{EL}$) is directly connected to the recombination rate of QD$^\text{X}$: $I_\text{EL} = k_{\text{X,r}} \cdot P_{\text{QD}^\text{X}}$.

In the steady condition of the single-dot EL cycle ($dP_\text{QD}/dt = dP_{\text{QD}^-}/dt = dP_{\text{QD}^\text{X}}/dt = 0$), we obtain:

$$k_\text{e} \cdot P_\text{QD} = k_\text{h}^- \cdot P_{\text{QD}^-} = k_{\text{X,r}} \cdot P_{\text{QD}^\text{X}} = I_\text{EL} \tag{10}$$

Accordingly, the injection-rate coefficients can be derived from the photon emission rate and the occupation probabilities of QD state and QD$^-$ state by the following equations:

$$k_\text{e} = \frac{I_\text{EL}}{P_\text{QD}} = \frac{1}{\tau_\text{QD}} \tag{11}$$

$$k_\text{h}^- = \frac{I_\text{EL}}{P_{\text{QD}^-}} = \frac{1}{\tau_{\text{QD}^-}} \tag{12}$$

where $\tau_{QD}$ and $\tau_{QD^-}$ are characteristic times of electron injection and hole injection, respectively, which can be regarded as the residence times of QD state and QD⁻ state.

Considering an overall detection efficiency of ~10%, $I_{EL}$ of a single QD under 2.1 V is estimated to be $1.33 \times 10^4$ s⁻¹. $P_{QD}$ and $P_{QD^-}$ are determined to be 65% and 35%, respectively. According to equation (11) and equation (12), these values lead to $k_e = 2.0 \times 10^4$ s⁻¹ ($\tau_{QD}$: 49 μs) and $k_h^- = 3.8 \times 10^4$ s⁻¹ ($\tau_{QD^-}$: 26 μs), respectively.

**Data Availability**

The data that support the finding of this study are available from the corresponding author upon reasonable request. The source data underlying Supplementary Figure 5e and Supplementary Figure 7 are provided as a Source Data file.

**References**

1. Brus, L. E. Electron-electron and electron-hole interactions in small semiconductor crystallites: the size dependence of the lowest excited electronic state. *J. Chem. Phys.* **80,** 4403–4409 (1984).

2. Murray, C. B., Norris, D. J. & Bawendi, M. G. Synthesis and characterization of nearly monodisperse CdE (E = sulfur, selenium, tellurium) semiconductor nanocrystallites. *J. Am. Chem. Soc.* **115,** 8706–8715 (1993).

3. Alivisatos, A. P. Semiconductor clusters, nanocrystals, and quantum dots. *Science* **271,** 933–937 (1996).

4. Protesescu, L. et al. Nanocrystals of cesium lead halide perovskites ($CsPbX_3$, X = Cl, Br, and I): novel optoelectronic materials showing bright emission with wide color gamut. *Nano Lett.* **15,** 3692–3696 (2015).

5. Pu, C. et al. Synthetic control of exciton behavior in colloidal quantum dots. *J. Am. Chem. Soc.* **139,** 3302–3311 (2017).

6. Colvin, V. L., Schlamp, M. C. & Alivisatos, A. P. Light-emitting-diodes made from cadmium selenide nanocrystals and a semiconducting polymer. *Nature* **370,** 354–357 (1994).

7. Coe, S., Woo, W. K., Bawendi, M. & Bulović, V. Electroluminescence from single monolayers of nanocrystals in molecular organic devices. *Nature* **420,** 800–803 (2002).

8. Sun, Q. et al. Bright, multicoloured light-emitting diodes based on quantum dots. *Nat. Photonics* **1,** 717–722 (2007).

9. Qian, L., Zheng, Y., Xue, J. & Holloway, P. H. Stable and efficient quantum-dot light-emitting diodes based on solution-processed multilayer structures. *Nat. Photonics* **5,** 543–548 (2011).

10. Mashford, B. S. et al. High-efficiency quantum-dot light-emitting devices with enhanced

charge injection. *Nat. Photonics* **7,** 407–412 (2013).

11. Dai, X. et al. Solution-processed, high-performance light-emitting diodes based on quantum dots. *Nature* **515,** 96–99 (2014).

12. Yang, Y. et al. High-efficiency light-emitting devices based on quantum dots with tailored nanostructures. *Nat. Photonics* **9,** 259–266 (2015).

13. Oh, N. et al. Double-heterojunction nanorod light-responsive LEDs for display applications. *Science* **355,** 616–619 (2017).

14. Li, X. et al. Bright colloidal quantum dot light-emitting diodes enabled by efficient chlorination. *Nat. Photonics* **12,** 159–164 (2018).

15. Cao, W. et al. Highly stable QLEDs with improved hole injection via quantum dot structure tailoring. *Nat. Commun.* **9,** 2608 (2018).

16. Lim, J. et al. Droop-free colloidal quantum dot light-emitting diodes. *Nano Lett.* **18,** 6645–6653 (2018).

17. Shen, H. et al. Visible quantum dot light-emitting diodes with simultaneous high brightness and efficiency. *Nat. Photonics* **13,** 192–197 (2019).

18. Sun, Y. et al. Investigation on thermally induced efficiency roll-off: toward efficient and ultrabright quantum-dot light-emitting diodes. *ACS Nano* **13,** 11433–11442 (2019).

19. Li, Y. et al. Stoichiometry-controlled InP-based quantum dots: synthesis, photoluminescence, and electroluminescence. *J. Am. Chem. Soc.* **141,** 6448–6452 (2019).

20. Won, Y.-H. et al. Highly efficient and stable InP/ZnSe/ZnS quantum dot light-emitting diodes. *Nature* **575,** 634–638 (2019).

21. Song, J. et al. Quantum dot light-emitting diodes based on inorganic perovskite cesium lead halides (CsPbX$_3$). *Adv. Mater.* **27,** 7162–7167 (2015).

22. Chiba, T. et al. Anion-exchange red perovskite quantum dots with ammonium iodine salts for highly efficient light-emitting devices. *Nat. Photonics* **12,** 681–687 (2018).

23. Bae, W. K. et al. Controlling the influence of Auger recombination on the performance of quantum-dot light-emitting diodes. *Nat. Commun.* **4,** 2661 (2013).

24. Shirasaki, Y., Supran, G. J., Tisdale, W. A. & Bulović, V. Origin of efficiency roll-off in colloidal quantum-dot light-emitting diodes. *Phys. Rev. Lett.* **110,** 217403 (2013).

25. Shendre, S., Sharma, V. K., Dang, C. & Demir, H. V. Exciton dynamics in colloidal quantum-dot LEDs under active device operations. *ACS Photonics* **5,** 480–486 (2018).

26. Schubert, E. F. *Light-Emitting Diodes* (Cambridge University Press, Cambridge, 2003).

27. Klimov, V. I. et al. Quantization of multiparticle Auger rates in semiconductor quantum dots. *Science* **287,** 1011–1013 (2000).

28. Jha, P. P. & Guyot-Sionnest, P. Trion decay in colloidal quantum dots. *ACS Nano* **3,** 1011–1015 (2009).

29. Lin, X. et al. Electrically-driven single-photon sources based on colloidal quantum dots with near-optimal antibunching at room temperature. *Nat. Commun.* **8,** 1132 (2017).

30. Kagan, C. R., Murray, C. B. & Bawendi, M. G. Long-range resonance transfer of electronic excitations in close-packed CdSe quantum-dot solids. *Phys. Rev. B* **54,** 8633–8643 (1996).

31. Ginger, D. S. & Greenham, N. C. Charge injection and transport in films of CdSe nanocrystals. *J. Appl. Phys.* **87,** 1361–1368 (2000).

32. Akselrod, G. M. et al. Subdiffusive exciton transport in quantum dot solids. *Nano Lett.* **14,** 3556–3562 (2014).

33. Park, Y. S., Bae, W. K., Pietryga, J. M. & Klimov, V. I. Auger recombination of biexcitons and negative and positive trions in individual quantum dots. *ACS Nano* **8,** 7288–7296 (2014).

34. Xu, W. et al. Deciphering charging status, absolute quantum efficiency, and absorption cross section of multicarrier states in single colloidal quantum dots. *Nano Lett.* **17,** 7487–7493 (2017).

35. Nirmal, M. et al. Fluorescence intermittency in single cadmium selenide nanocrystals. *Nature* **383,** 802–804 (1996).

36. Meng, R. et al. Charging and discharging channels in photoluminescence intermittency of single colloidal CdSe/CdS core/shell quantum dot. *J. Phys. Chem. Lett.* **7,** 5176–5182 (2016).

37. Galland, C. et al. Two types of luminescence blinking revealed by spectroelectrochemistry of single quantum dots. *Nature* **479,** 203–U275 (2011).

38. Alperson, B., Cohen, S., Rubinstein, I. & Hodes, G. Room-temperature conductance spectroscopy of CdSe quantum dots using a modified scanning force microscope. *Phys. Rev. B* **52,** R17017–R17020 (1995).

39. Zabet-Khosousi, A. & Dhirani, A.-A. Charge transport in nanoparticle assemblies. *Chem. Rev.* **108,** 4072–4124 (2008).

40. Talapin, D. V., Lee, J.-S., Kovalenko, M. V. & Shevchenko, E. V. Prospects of colloidal nanocrystals for electronic and optoelectronic applications. *Chem. Rev.* **110,** 389–458 (2010).

41. Bozyigit, D., Yarema, O. & Wood, V. Origins of low quantum efficiencies in quantum dot LEDs. *Adv. Funct. Mater.* **23,** 3024–3029 (2013).

42. Fan, F. et al. Continuous-wave lasing in colloidal quantum dot solids enabled by facet-selective epitaxy. *Nature* **544,** 75–79 (2017).

43. Lim, J., Park, Y.-S. & Klimov, V. I. Optical gain in colloidal quantum dots achieved with direct-current electrical pumping. *Nat. Mater.* **17,** 42–49 (2017).

44. Zhang, Z. et al. High-performance, solution-processed, and insulating-layer-free light-

emitting diodes based on colloidal quantum dots. *Adv. Mater.* **30,** 1801387 (2018).

45. Lakowicz, J. R. *Principles of Fluorescence Spectroscopy* (Springer, New York, 2006).

**Acknowledgments**

We thank Mr. Yang Liu for assistance with in-situ measurements of PL intensity of the QD film in working QLEDs and Mr. Zhenlei Yao for fabricating the high-efficiency QLEDs. We also thank Prof. Jianpu Wang and Prof. Feng Gao for valuable advice on preparing the manuscript. This work was financially supported by the National Key Research and Development Program of China (2016YFB0401600, 2018YFB2200401), the National Natural Science Foundation of China (51522209, 21975220, 91833303, 51911530155, 91733302, 61635009 and 61975180), the Fundamental Research Funds for the Central Universities (2017XZZX001-03A, 2019QNA5005), and Zhejiang University Education Foundation Global Partnership Fund.

**Author contributions**

Y.J. supervised the work and finalized the manuscript. Y.J. and Y.D. conceived the idea and designed the experiments. Y.D. fabricated the single-dot EL devices, carried out the electrically-pumped single-nanocrystal spectroscopy measurements, the in-situ/operando spectroscopy measurements on QD-LEDs and analysed the results. X.L. designed and built the setup for single-nanocrystal measurements and contributed to the analyses of spectroscopic results. X.P., W.F., L.W., D.D. and R.H.F. participated in data analysis and provided major revisions. All authors discussed the results and commented on the manuscript.

**Competing interests**

The authors declare no competing interests.

**Additional information**

Correspondence and requests for materials should be addressed to Y.J (yizhengjin@zju.edu.cn).

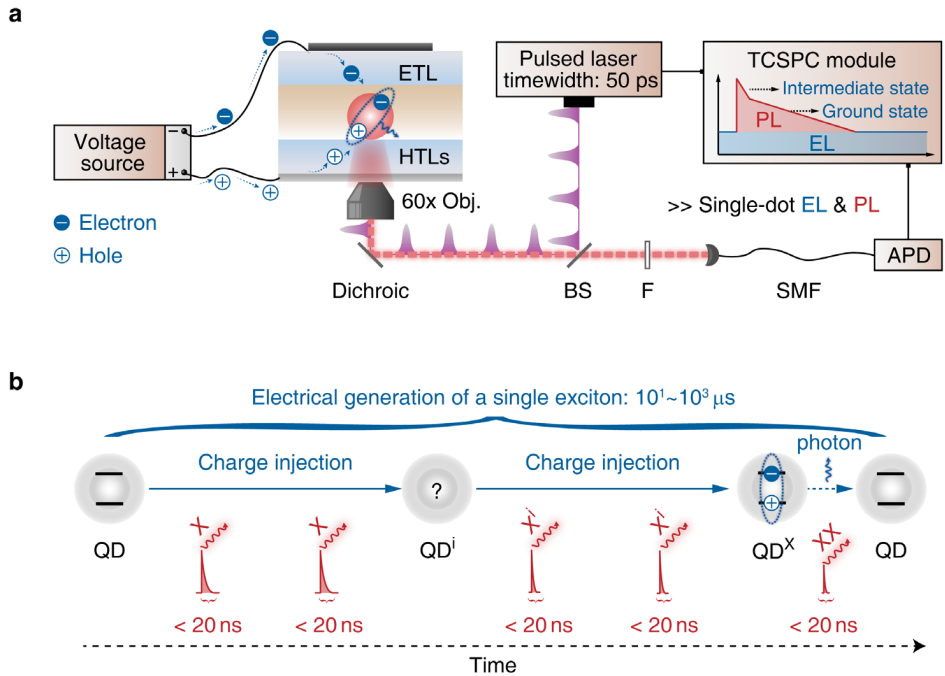

**Fig. 1: Electrically-pumped single-nanocrystal spectroscopy.**

**a** Schematic diagram of the experimental setup. Electrical excitation (voltage source) and optical excitation (450 nm pulsed laser) are simultaneously applied on the single QD. Intensities and time-correlated single-photon counting data are recorded simultaneously, enabling probe of states relevant to the single-dot EL. Optical excitation is kept in the single-exciton regime (average excitation per pulse < 0.1) so that it does not alter the charging status of the single QD. **b** Temporal evolution of a single QD under electrical excitation. The pulsed laser serves as a fast and non-invasive probe. The average time of an EL cycle is substantially longer than the characteristic lifetimes of optically-excited states. QD$^i$ represents the possible intermediate states for single-dot EL processes (blue arrows). The corresponding optically-excited states are denoted as X$^i$ (in red).

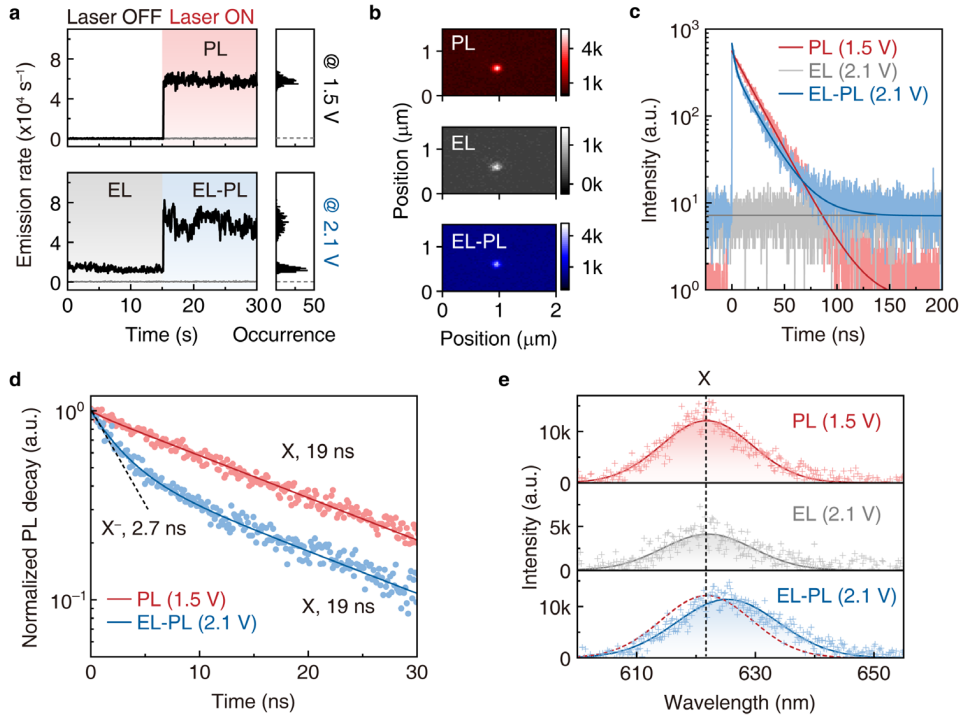

**Fig. 2: QD⁻ as the intermediate state for the electrical generation of a single exciton.**

**a** Intensity-time traces of a single QD under different excitation conditions. "PL" (red shaded regions), "EL" (grey shaded regions) and "EL-PL" (blue shaded regions) traces correspond to emissions recorded under optical excitation, electrical excitation, and simultaneous optical and electrical excitation, respectively. Grey lines are background counts. **b** Microscopy images of the single QD under different excitation conditions. **c** TCSPC data of PL (red), EL (grey) and EL-PL (blue) emissions. The data of EL emission (grey), which matches with the baseline of the EL-PL emission, is also shown. **d** PL decay for the single QD with electrical injection (blue) or without electrical injection (red), extracted from **c**. Solid lines are exponential fits. **e** Single-dot spectra measured under different conditions. The single-exciton PL spectrum (red-dashed line) is also shown at the bottom for comparison.

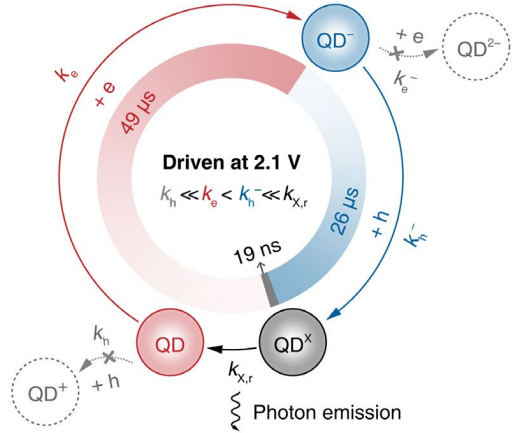

**Fig. 3: Dynamics of the single-dot EL cycle.**

EL of a single QD at 2.1 V consists of three elementary steps, i.e., electron injection into the neutral QD (red arrow, rate coefficient: $k_e = 2.0 \times 10^4$ s$^{-1}$), hole injection into the negatively charged QD (blue arrow, rate coefficient: $k_h^- = 3.8 \times 10^4$ s$^{-1}$) and radiative recombination of an exciton (black arrow, rate coefficient: $k_{X,r} = 1/19$ ns$^{-1}$). At 2.1 V, hole injection into the neutral QD (QD $\xrightarrow{+h}$ QD$^+$) or electron injection into the negatively charged QD (QD$^- \xrightarrow{+e}$ QD$^{2-}$) is negligible according to spectroscopic results. Residence times of the three states, i.e., characteristic times of three elementary processes, are illustrated, indicating a long-lived intermediate state of QD$^-$.

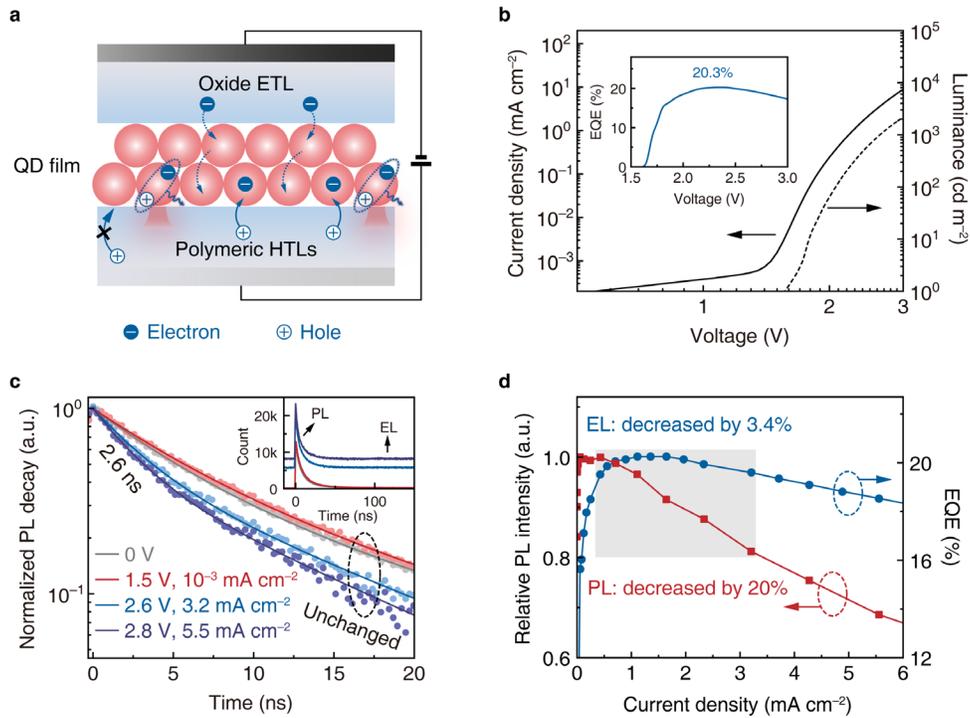

**Fig. 4: Exciton generation in a high-efficiency QD-LED.**

**a** schematic diagram showing single-exciton generation via $QD^-$ ($QD \xrightarrow{+e} QD^- \xrightarrow{+h} QD^X$) in the QD-LED. Holes are injected into the negatively-charged QDs close to the interface of hole-transport layer/QD layer. **b** Current density and luminance versus voltage characteristics. Inset: the EQE-voltage curve. **c** PL decay of the QD film in the working QD-LED driven at different voltages, measured by the microscopy system shown in Supplementary Figure 11. Inset: the original TCSPC data showing baselines originated from EL emissions. In the experiments, photon counting rates are reduced by using an extra neutral density filter to meet the requirement of TCSPC technique. **d** EQE and relative PL intensity of the QD-LED at different current densities. The grey-shaded region corresponds to the high-efficiency regime, in which a significant discrepancy between changes in the EL and PL efficiencies is observed.

# Supplementary Information For

**Deciphering exciton-generation processes in quantum-dot electroluminescence**

Deng *et al.*

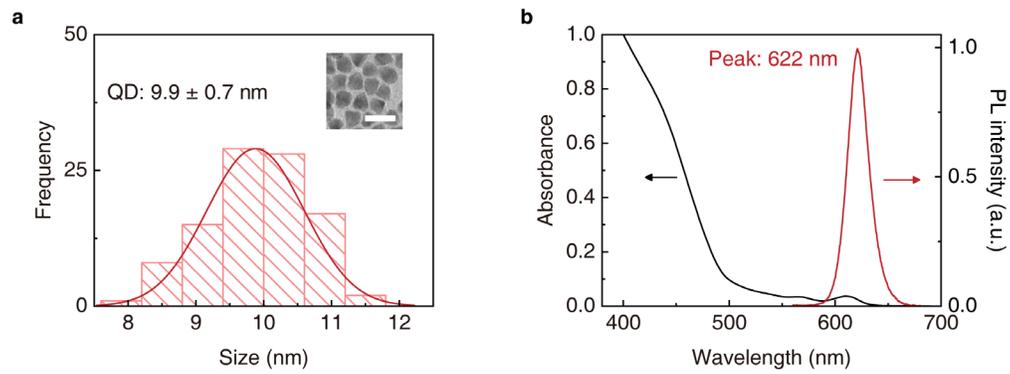

**Supplementary Figure 1: The CdSe-CdZnS QDs. a** Size distribution histogram of the CdSe-CdZnS QDs. Inset: a typical TEM image (scale bar: 20 nm). **b** Ultraviolet-visible absorption and PL spectra of the solution of QDs in octane.

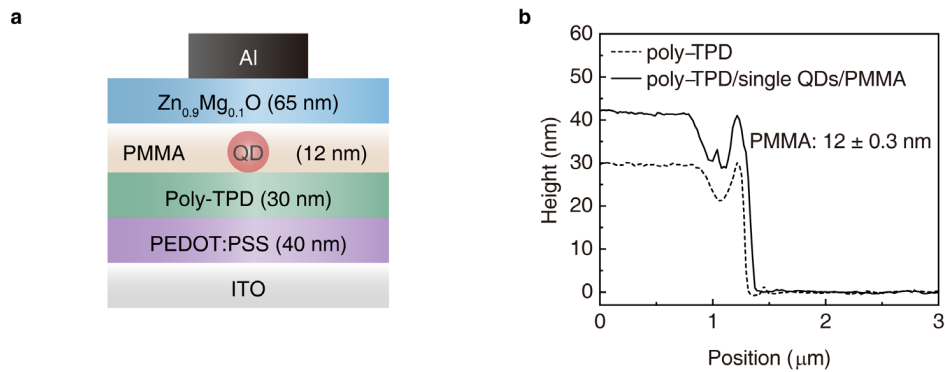

**Supplementary Figure 2: A single-dot EL device. a** Device structure: ITO/PEDOT:PSS/poly-TPD/Single QD in PMMA/Zn$_{0.9}$Mg$_{0.1}$O/Al. The isolated QDs are deposited onto the polymeric hole-transport layers and encapsulated by a thin layer of PMMA. The Zn$_{0.9}$Mg$_{0.1}$O nanocrystals are used as an electron-transport layer. **b** Thickness measurement of the PMMA layer. Cross-section profiles are measured by an atomic force microscopy in the tapping mode. Results demonstrate a flat surface (solid line, position: 0–0.7 μm) after the deposition of the sparsely dispersed single QDs and the PMMA layer onto the poly-TPD film. The overall thickness of the PMMA layer (12 ± 0.3 nm) is slightly larger than the average diameter of the CdSe-CdZnS QDs (9.9 ± 0.7 nm), suggesting that the single QDs are immersed into the PMMA layer (as illustrated in the device structure). Note that the trenches (position: 0.8–1.3 μm) are caused by scratching procedures to create the steps required for the thickness measurements.

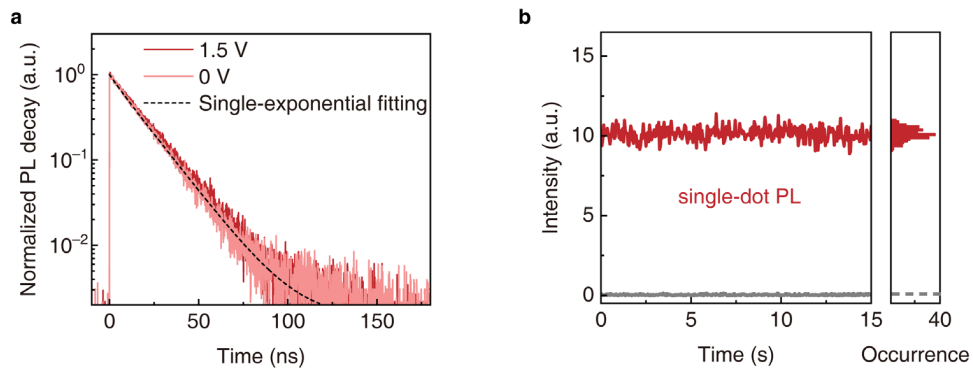

**Supplementary Figure 3: PL properties of a single QD in the single-dot EL device. a** PL decay curves of a single QD measured at 0 V and 1.5 V (below the turn-on threshold, ~1.85 V). The two curves are nearly identical and both curves can be fitted by a single-exponential function with a characteristic lifetime of 19 ns (dashed line). The results suggest negligible electric-field-induced effects for the QD in the single-dot EL device. Background PL emission from the hole-transport layer (poly-TPD) in the device (contribution < 5%) was subtracted. **b** PL intensity-trace of a single QD in the device and the corresponding histogram. The grey lines are background baselines. The non-blinking feature, i.e., no switching between states with different PL efficiencies, suggests that photo-induced ionization processes are negligible. These properties indicate that the single QD in the single-dot EL device offers stable single-exciton emission under our optical-excitation conditions.

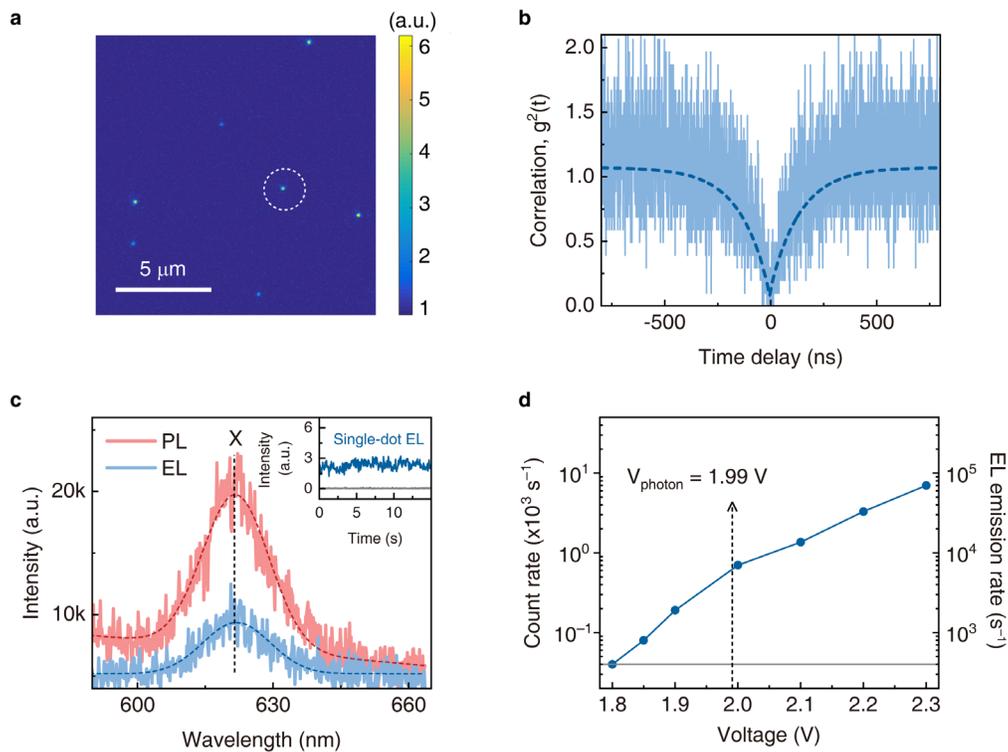

**Supplementary Figure 4: EL behaviours of the single-dot device. a** Microscopic image of EL from isolated single QDs (driven at 2.15 V). Scale bar: 5 μm. **b** A second-order correlation function ($g^2(t)$, blue line) of single-dot EL (driven at 2.15 V). The value of $g^2(0)$ is determined to be 0.07 by exponential fitting (blue dashed line), indicating minimal bi-exciton emission in the single-dot EL[1, 2]. **c** PL (red) and EL (blue, driven at 2.1 V) spectra of a single QD share the same peak wavelength of 622 nm and an identical full width at half maximum (FWHM) of 18 nm (integration time: 4 s). Inset: An EL intensity-time trace of the single QD showing stable single-exciton emission. The emission on the short wavelength side of the PL spectrum originates from PL of the hole-transporting layer, poly-TPD. The absence of background emission from poly-TPD in the EL spectrum suggests the negligible leakage current. Dashed lines are fits to the data. **d** Voltage-dependent EL intensity of a single QD (circled in **a**). The internal EL emission rates (right) are derived from the photon-counting rates (left) according to an overall detection efficiency of ~10%. The background counts are illustrated by the grey line. The threshold voltage for the single-dot EL is ~1.85 V, which is below the voltage (dashed black arrow) corresponding to the bandgap of the QD (1.99 eV).

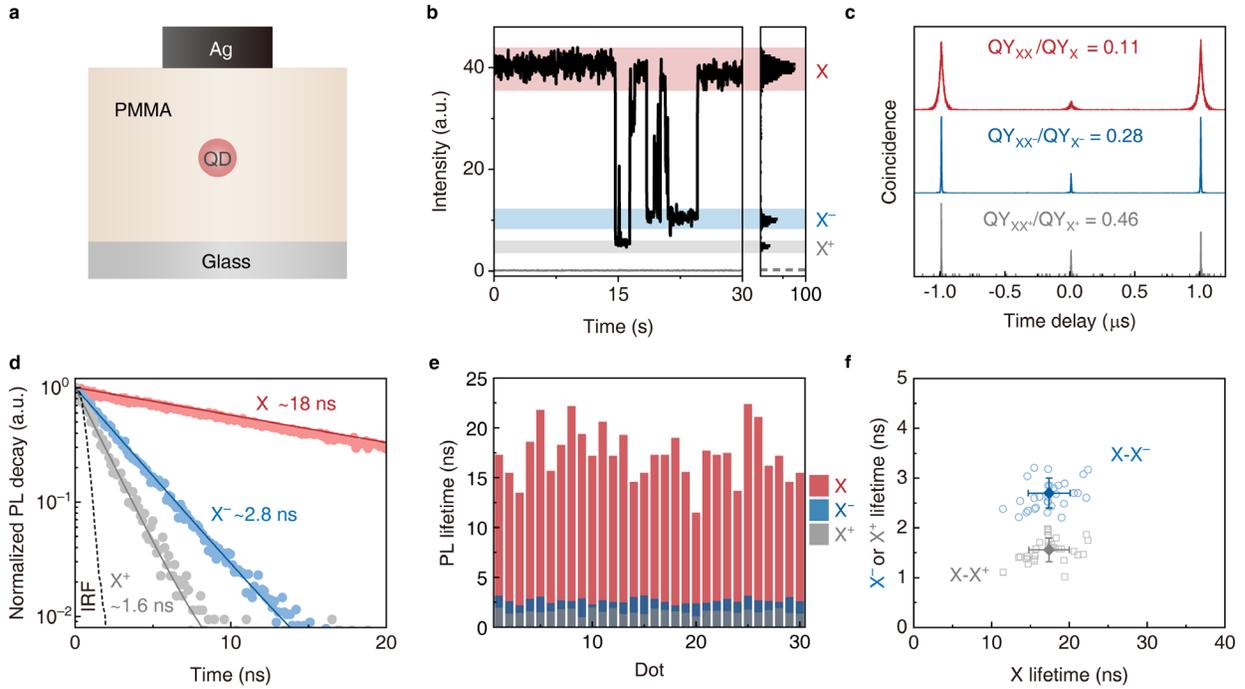

**Supplementary Figure 5: Optical properties of X, $X^-$ and $X^+$ of a single CdSe-CdZnS QD in the presence of a metal electrode. a** Single QDs were dispersed in a PMMA matrix (thickness ~100 nm) by spin-coating a solution of QDs and PMMA in toluene (3:100 by weight) onto a glass slide, followed by deposition of a silver electrode. **b** Representative PL intensity-time trace of a single QD (measured under pulsed laser excitation) and the corresponding histogram showing switching between three emissive states. For CdSe-based QDs, the "bright state" (red shaded), the "dim state" with a higher QY (blue shaded) and the "dim state" with a lower QY (grey shaded) are assigned as X, $X^-$ and $X^+$, respectively. $g^2(t)$ curves and PL decay curves for the three states can be selectively extracted from the corresponding intensity-windows. **c** $g^2(t)$ curves for X (red), $X^-$ (blue) and $X^+$ (grey) of the single QD extracted according to the temporal intensities. The PL QY ratios of bi-exciton to single exciton are derived by following an established method[3]. According to a recent report[4], $QY_{XX^-}/QY_{X^-}$ is smaller than $QY_{XX^+}/QY_{X^+}$ for CdSe-based QDs because of the different radiative and Auger recombination pathways of $X^-$ and $X^+$. Thus, the results confirm our assignments of $X^-$ and $X^+$. **d** PL decay curves of for X (red), $X^-$ (blue) and $X^+$ (grey) of the single QD extracted according to the temporal intensities. The solid lines are the single-exponential fitted curves. The grey dashed curve represents the instrument response function (0.78 ns). **e** PL lifetimes of X (red), $X^-$ (blue) and $X^+$ (black) of 30 single QDs. **f** The corresponding statistical

distribution of PL lifetimes. Blue squares and grey circles represent lifetime correlations of X-X$^-$ and X-X$^+$ for each QD, respectively. Average values and the corresponding standard deviations of the PL lifetimes are illustrated by diamonds with error bars. The characteristic lifetimes of X, X$^-$ and X$^+$ are determined to be 17.4 ± 2.6 ns, 2.7 ± 0.3 ns, and 1.6 ± 0.2 ns, respectively. We note that in the single-dot EL device, the lifetime of X is decreased due to the electrode-induced Purcell effects[5]. Here, the photonic environment of the single QDs in the single-dot EL device is mimicked by the structure shown in **a**. The average X lifetime (17.4 ns) for the QDs agrees well with that in single-dot EL devices (17–19 ns), validating the similarity of the two photonic environments. Therefore, statistical results for X$^-$ and X$^+$ displayed in **f** are used to identify charged states of single QDs in EL devices. The source data of lifetime distributions are provided as a Source Data file.

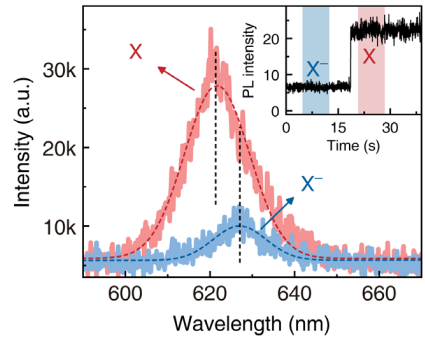

**Supplementary Figure 6: PL spectra of X (red) and X⁻ (blue) of a single CdSe-CdZnS QD.** The QD was incorporated into the structure shown in Supplementary Figure 5a. PL from a single QD was split by a 50:50 beam splitter and collected by two optical fibres. One output was sent to a spectrometer to record the single QD spectra, and another output was detected by an APD to monitor temporal intensities (inset). In this way, during the intensity blinking of a single QD, spectra of long "bright" durations corresponding to X emission, and spectra of long "dim" durations corresponding to X⁻ emission (PL intensity: ~25 % of "bright" states) were selectively recorded (integration time: 4 s). The peak of X⁻ emission is red-shifted by 5 nm in the wavelength (16 meV in photon energy) comparing with that of X emission. Dashed lines are fits to the data.

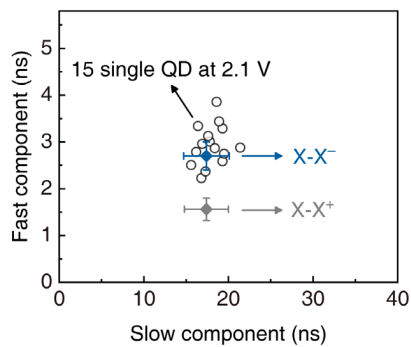

**Supplementary Figure 7: Electrically-pumped single-nanocrystal spectroscopy on 15 different single QDs.** Lifetimes (circles) of slow components (18.0 ± 1.5 ns) and fast components (2.9 ± 0.4 ns) are in agreement with characteristic lifetimes of X and $X^-$ (blue diamond with the error bar, obtained from Supplementary Figure 5). The significant deviation from the characteristic lifetime of $X^+$ (grey diamond with the error bar, obtained from Supplementary Figure 5) excludes the possibility of $QD^+$ as the intermediate state. Error bars represent the corresponding standard deviations. The source data of lifetime distributions are provided as a Source Data file.

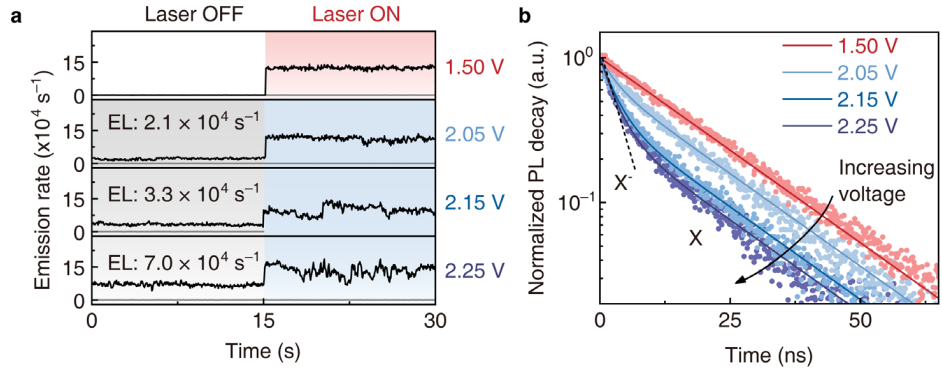

**Supplementary Figure 8: Electrically-pumped single-nanocrystal spectroscopy on a single QD under different electrical-injection levels. a** Intensity-time traces of a single QD under optical excitation (red shaded region), electrical excitation (grey shaded regions) and simultaneous optical and electrical excitation (blue shaded regions). The EL emission rates increase at higher voltages, leading to stable emission of up to ~$7\times10^4$ s$^{-1}$. **b** Normalized PL decay curves of the single QD under different voltages (blue lines). PL decay of the single QD without electrical injection (red lines, measured at 1.50 V) is shown for comparison. All curves can be well fitted by double-exponential functions with fixed components of X$^-$ (2.9 ns) and X (17 ns), as are shown by the solid lines. The fractional contributions of X$^-$ become greater as the electrical injection level increases. Dashed lines are fits to the data.

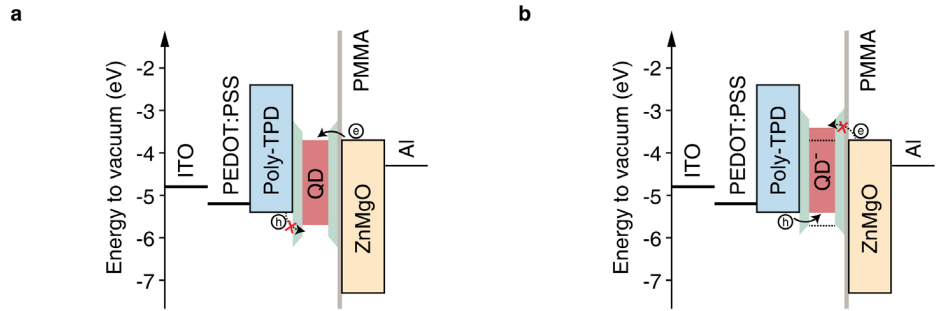

**Supplementary Figure 9: Energy-level diagram of a single-dot EL device. a** Schematic diagram for charge injection into a neutral QD. From an energetical point of view, electron injection (solid arrow) is favoured while hole injection (dashed arrow) is hindered due to the considerable offset between HOMO level of poly-TPD and the valance-band level of the CdSe-CdZnS QD. **b** Schematic diagram for charge injection into a negatively-charged QD. Due to Coulomb charging effects, the addition of one electron can shift the electrical potential of the single QD toward higher energy[6]. According to a classic electrostatic model[7], the upward energy shift ($\Delta E_c$), i.e., the single-electron charging energy, is calculated from the capacitance of the QD (C) by the equation of $\Delta E_c = e^2/2C = e^2/(8\pi\varepsilon_0\varepsilon_r r)$. In the equation, $e$, $r$, and $\varepsilon_r$ are the elementary electronic charge, the core radius of the QD, and the relative dielectric constant of the surrounding medium, respectively. For our QDs with a core radius of ~1.6 nm, the energy shift is estimated to be ~150 meV. This value agrees reasonably well with the charging energy measured from CdSe QDs with similar diameters[8]. Consequently, the hole-injection energy barrier for hole injection is reduced and the electron-injection energy barrier is increased.

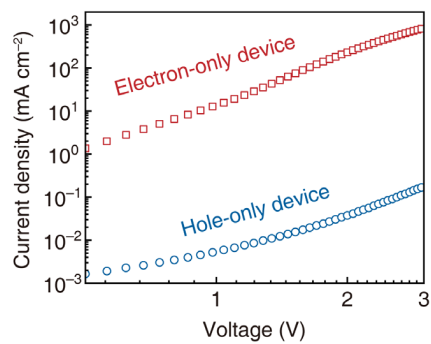

**Supplementary Figure 10: Current densities versus voltages of an electron-only device and a hole-only device.** The structure of the electron-only device is ITO/QDs (25 nm)/$Zn_{0.9}Mg_{0.1}O$ (65 nm)/Ag. The structure of the hole-only device is ITO/PEDOT:PSS (40 nm)/TFB (45 nm)/QDs (25 nm)/$MoO_3$ (5 nm)/Au.

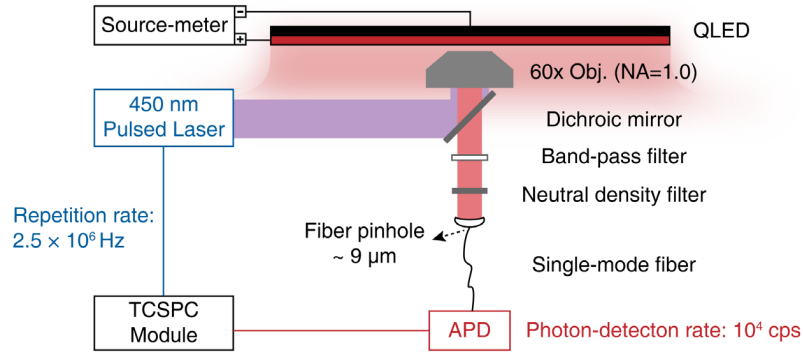

**Supplementary Figure 11: Transient PL measurements on the working QD-LED.** The experimental setup shares the same microscopic system shown in Fig. 1. The QD film was under simultaneous electrical excitation by a DC bias and optical excitation by a pulsed laser (450 nm, 2.5 MHz). An optical fibre (core size: ~9 μm) was used to collect the emissions from a micro area of the device. PL decay curves are extracted by subtracting the EL emissions from the micro area of the working QD-LED. We highlight that an extra neutral density filter is placed in front of the single-mode fibre so that the requirement of TCSPC measurement (the photon-counting rate is two orders of magnitude smaller than the repetition rate of the pulsed laser) is satisfied[9]. In addition, both optical and electrical excitation are controlled in the single-exciton regime (see the Methods section for details of the excitation conditions) to ensure that optical excitation and electrical excitation are independent of each other.

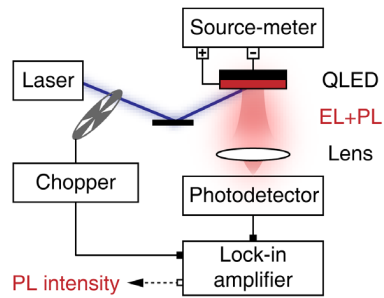

**Supplementary Figure 12: In-situ measurements of relative PL intensity of the QD film in the working QD-LED.** A frequency-modulated (1,003 Hz) low-intensity optical excitation (405 nm) is applied to the operating QD-LED. The relative PL intensities of the QD film are resolved by a lock-in amplifier.

**Supplementary References**

1. Lounis, B. et al. Photon antibunching in single CdSe/ZnS quantum dot fluorescence. *Chem. Phys. Lett.* **329,** 399–404 (2000).
2. Lin, X. et al. Electrically-driven single-photon sources based on colloidal quantum dots with near-optimal antibunching at room temperature. *Nat. Commun.* **8,** 1132 (2017).
3. Nair, G., Zhao, J. & Bawendi, M. G. Biexciton quantum yield of single semiconductor nanocrystals from photon statistics. *Nano Lett.* **11,** 1136–1140 (2011).
4. Xu, W. et al. Deciphering charging status, absolute quantum efficiency, and absorption cross section of multicarrier states in single colloidal quantum dots. *Nano Lett.* **17,** 7487–7493 (2017).
5. Barnes, W. L. Fluorescence near interfaces: the role of photonic mode density. *J. Mod. Opt.* **45,** 661–699 (1998).
6. Talapin, D. V., Lee, J.-S., Kovalenko, M. V. & Shevchenko, E. V. Prospects of colloidal nanocrystals for electronic and optoelectronic applications. *Chem. Rev.* **110,** 389–458 (2010).
7. Zabet-Khosousi, A. & Dhirani, A.-A. Charge transport in nanoparticle assemblies. *Chem. Rev.* **108,** 4072–4124 (2008).
8. Alperson, B., Cohen, S., Rubinstein, I. & Hodes, G. Room-temperature conductance spectroscopy of CdSe quantum dots using a modified scanning force microscope. *Phys. Rev. B* **52,** R17017–R17020 (1995).
9. Lakowicz, J. R. *Principles of Fluorescence Spectroscopy* (Springer, New York, 2006).